# MgB$_2$ coated superconducting tapes with high critical current densities fabricated by hybrid physical–chemical vapor deposition

**Mahipal Ranot, W. N. Kang***

*BK21 Physics Division and Department of Physics, Sungkyunkwan University, Suwon 440-746, Republic of Korea*

**Abstract**

The MgB$_2$ coated superconducting tapes have been fabricated on textured Cu (0 0 1) and polycrystalline Hastelloy tapes using coated conductor technique, which has been developed for the second generation high temperature superconducting wires. The MgB$_2$/Cu tapes were fabricated over a wide temperature range of 460–520 °C by using hybrid physical–chemical vapor deposition (HPCVD) technique. The tapes exhibited the critical temperatures ($T_c$) ranging between 36 and 38 K with superconducting transition width ($\Delta T_c$) of about 0.3–0.6 K. The highest critical current density ($J_c$) of $1.34 \times 10^5$ A/cm$^2$ at 5 K under 3 T is obtained for the MgB$_2$/Cu tape grown at 460 ºC. To further improve the flux pinning property of MgB$_2$ tapes, SiC is coated as an impurity layer on the Cu tape. In contrast to pure MgB$_2$/Cu tapes, the MgB$_2$ on SiC-coated Cu tapes exhibited opposite trend in the dependence of $J_c$ with growth temperature. The improved flux pinning by the additional defects created by SiC-impurity layer along with the MgB$_2$ grain boundaries lead to strong improvement in $J_c$ for the MgB$_2$/SiC/Cu tapes. The MgB$_2$/Hastelloy superconducting tapes fabricated at a temperature of 520 °C showed the critical temperatures ranging between 38.5 and 39.6 K. We obtained much higher $J_c$ values over the wide field range for MgB$_2$/Hastelloy tapes than the previously reported data on other metallic substrates, such as Cu, SS, and Nb. The $J_c$ values of $J_c$(20 K, 0 T) ~$5.8 \times 10^6$ A/cm$^2$ and $J_c$(20 K, 1.5 T) ~$2.4 \times 10^5$ A/cm$^2$ is obtained for the 2-µm-thick MgB$_2$/Hastelloy tape. This paper will review the merits of coated conductor approach along with the HPCVD technique to fabricate MgB$_2$ conductors with high $T_c$ and $J_c$ values which are useful for large scale applications.

* Corresponding author.
Prof. Won Nam Kang
Postal address: Department of Physics, Sungkyunkwan University, Suwon 440-746, Republic of Korea
Phone: +82-31-290-5904
Fax: +82-31-290-7055
E-mail address: wnkang@skku.edu

**Contents**

## 1. Introduction

The exciting magnesium diboride ($MgB_2$) superconductor is a promising material for superconducting magnets and microelectronic devices at low cost because of its remarkably high critical temperature ($T_c$) of 39 K [1,2], which is 2–4 times higher than those of conventional metallic superconductors, such as Nb-Ti ($T_c$ = 9 K) and $Nb_3Sn$ ($T_c$ = 18 K). The strongly linked nature of the intergrains [3] with a high charge carrier density [4] in this material makes it further an attractive candidate for the next generation of superconductor applications. The practical application of any superconductor strongly depends on its capacity for carrying loss-free currents in high magnetic fields [5]. Therefore, $MgB_2$ superconductors in the form of wires and tapes with high critical current density ($J_c$) are required to be produced in long lengths to replace the conventional metallic superconductors [6,7]. Enormous efforts have been devoted to the fabrication and characterization of $MgB_2$ conductors, such as wires and tapes by the powder-in-tube (PIT) process with encouraging results [8–10]. Already some companies have commercially produced $MgB_2$ wires from conventional powder by using PIT method [11,12]. But the PIT-processed conductors are facing the problems of poor self-field and in-field $J_c$ values of the order of $10^5$–$10^6$ $A/cm^2$ and low packing density of $MgB_2$ (below ~50%) [13,14]. In contrast to PIT bulks and conductors, there are several reports on $MgB_2$ thin films with very high upper critical field ($H_{c2}$), high self-field $J_c$ of ~$10^7$ $A/cm^2$ and high $J_c$ in magnetic fields [15–17]. These high $J_c$ values obtained in these films formed by vapor deposition techniques demonstrate the potential for further improving the current carrying capabilities of $MgB_2$ wires and tapes. However, there are only few reports on the $MgB_2$ conductors made by $MgB_2$ film fabrication processes [18–23].

In this article, we report our recent systematic investigation on the fabrication of $MgB_2$ coated superconducting tapes by depositing $MgB_2$ films on textured Cu (0 0 1) and polycrystalline Hastelloy tapes by hybrid physical–chemical vapor deposition (HPCVD) technique. First, we will briefly describe the techniques used for the fabrication of $MgB_2$ wires and tapes. Then we describe the HPCVD technique used for the growth of $MgB_2$ coated conductors. The effect of growth temperature on the microstructure and superconducting properties of $MgB_2$/Cu (0 0 1) superconducting tapes is investigated and discussed. Then we will show how the tape performances can be improved by creating additional defects in the $MgB_2$ films by SiC-coated Cu tapes. The results on $MgB_2$/Hastelloy superconducting tapes are also presented. Finally, we will conclude our results.

## 2. Methods for the fabrication of $MgB_2$ conductors

A number of techniques have been developed to fabricate $MgB_2$ wires and tapes with high critical current densities. The following section describes those methods briefly.

### 2.1. Diffusion method

Soon after the discovery of superconductivity in $MgB_2$, the first $MgB_2$ wires fabrication is demonstrated by Canfield et al. by exposing boron filaments to Mg vapor [24]. These wires showed exceptionally low resistivity, high density and high $T_c$ of 39.4 K. The $J_c$ of the order of $5 \times 10^5$ $A/cm^2$ at 5 K and self-field is obtained for these wires. DeFouw et al. report the synthesis of superconducting $MgB_2$ fibers within a magnesium matrix by infiltration of a B fiber preform with liquid Mg and subsequent *in situ* reaction at 950 °C [25]. The $MgB_2$ fibers exhibit a critical temperature slightly above 39 K and a critical current density of 360 $kA/cm^2$ at 5 K and self-field. Diffusion method was also used for the synthesis of $MgB_2$ tapes. Togano et al. fabricated a B/Fe–Mg composite tape and obtained a high-density $MgB_2$ layer by the diffusion reaction between the B plate and a Fe–Mg alloy sheet [26].

### 2.2. Powder-in-tube process

The powder-in-tube (PIT) process is one of the most commonly used methods for the fabrication of $MgB_2$ wires and tapes. The PIT process is classified into two different processes: *in situ* and *ex situ*. The former employs powder starting materials [27], whereas the latter employs a powder of synthesized superconducting material [28]. The superconducting properties of PIT-processed $MgB_2$ conductors are very sensitive to the quality of the starting powder [29–33], particle size [34–36], porosity of the $MgB_2$ core [37,38], heat treatment temperature [34,39–41], and the sheath material [27,42]. In general, *ex situ*-processed $MgB_2$ tapes show lower $J_c$ than *in situ*-processed tapes. Recently, Nakane et al. succeeded in the fabrication of high $J_c$ *ex situ* $MgB_2$/Fe tapes using $MgB_2$ cores removed from *in situ* $MgB_2$ tapes [43]. The key factor to obtain a high $J_c$ for an *ex situ* tape is the high quality of the $MgB_2$ starting powder. The highest $J_c$ values obtained so far at 4.2 K and 10 T is 27 kA/cm$^2$ and at 20 K and 5 T is 10 kA/cm$^2$.

Great progress has been made [44,45] and is going on for the fabrication of $MgB_2$ wires and tapes by PIT process to demonstrate the potential of $MgB_2$ superconductor in practical applications. However, the main problem of the PIT-processed conductors is the low packing density of $MgB_2$ (below ~50%). A further improvement in current carrying capability could be expected if the packing density of $MgB_2$ conductors can be increased.

### 2.3. Coated conductor technique

The coated conductor technique was basically developed for the cuprates high-$T_c$ superconductors. For example, $YBa_2Cu_3O_{7-\delta}$ (Y-123) compound, where Y-123 thin films are deposited on various buffer layers on a metallic substrate [46], the highly textured Y-123 is required for high critical current density in Y-123 coated conductor due to the poor superconducting coupling across the grain boundaries of this material [47]. A schematic representation of coated conductor technique is shown in Fig. 1a [48].

The fabrication of $MgB_2$ superconducting films on flexible metallic substrates using the coated conductor approach was firstly demonstrated by Komori et al. [18]. First, they prepared 1 μm thick yttria-stabilized-zirconia (YSZ) buffer layer on Hastelloy (C-276) tapes by bias sputtering. Subsequently, $MgB_2$ films were deposited on YSZ coated Hastelloy tapes using a KrF excimer laser with 400 mJ/pulse operating at 5 Hz. $MgB_2$ superconducting tape showed very high in-field transport critical current density of the order of $1.1 \times 10^5$ A/cm$^2$ at 4.2 K and 10 T. Ferrando et al. succeeded in fabricating carbon-alloyed $MgB_2$ coated conductors on round SiC fibers with a tungsten core by using HPCVD technique [21]. The diameter of the SiC fiber is 100 μm with a tungsten core of 10 μm. The carbon-alloyed fibers show a high upper critical field of 55 T at 1.5 K and a high irreversibility field of 40 T at 1.5 K. From economical and practical application point of view, the SiC fiber used in this work is not the desirable substrate for making wires because it is brittle and costly. $MgB_2$ coatings can also be made on inexpensive and strong metallic wires and tapes such as stainless steel and copper.

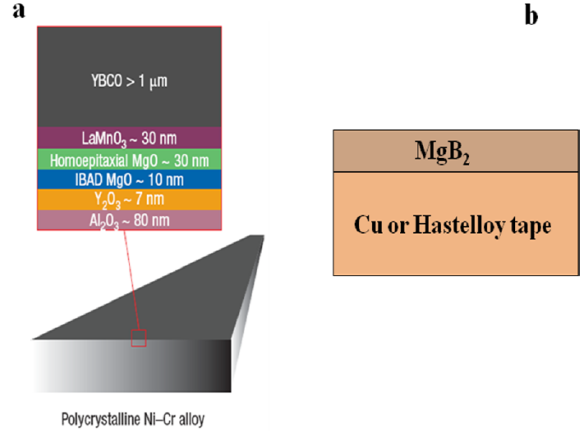

Fig. 1. (a) A schematic representation of coated conductor technique developed for the second generation high temperature superconductors. (b) A schematic we used for the fabrication of $MgB_2$ coated conductors.

Abe et al. reported the growth of $MgB_2$ films on metallic stainless steel substrates using the molten-salts electroplating (MSEP) technique [22]. Transport measurements have shown that the electroplated $MgB_2$ films possess an upper critical field, $H_{c2}(0)$ of 28 T, an irreversibility field, $H_{irr}(0)$ of 13 T, and a critical current density, $J_c$(5 K, 0 T) of 25000 A/cm$^2$, which are comparable with the PIT-processed $MgB_2$ tapes. The $MgB_2$ thin films deposited by electron beam evaporation technique on polycrystalline Cu plates by Masuda et al. [23] showed high $J_c$ of $1 \times 10^7$ A/cm$^2$ at 4.2 K and 1 T.

These results for MgB$_2$ with high upper critical field and high critical current densities demonstrate great potential of MgB$_2$ coated conductors for superconducting magnets. MgB$_2$ coated conductors also promise lower cost [49].

## 3. HPCVD for MgB$_2$ coated conductors

The film growth of MgB$_2$ superconductors has been achieved using molecular beam epitaxy (MBE) [50–52], pulsed laser deposition (PLD) [2,53–55], hybrid physical–chemical vapor deposition (HPCVD) [56–58], electron beam evaporation (EBE) [59–61], and the modified reactive method [62,63] etc. Among all these techniques, HPCVD has been proved to be the most effective one for the fabrication of high-quality MgB$_2$ films with superior superconducting properties [58,64]. For MgB$_2$ film growth, HPCVD process uses diborane (B$_2$H$_6$) as the boron precursor gas, but unlike conventional CVD, which only uses gaseous sources, heated bulk magnesium pieces are used as the Mg source in the deposition process. Since the process involves chemical decomposition of precursor gas and physical evaporation of metal bulk, it is named as hybrid physical-chemical vapor deposition.

Our HPCVD system [57], which is similar to that reported by Zeng et al. [56], consists of a vertical quartz tube reactor having a bell shape (inner diameter: 65 mm, height: 200 mm), gas inlet and flow control system, pressure maintenance system, temperature control system, gas exhaust and cleaning system, inductively coupled heater, and a load-lock chamber. Fig. 2a shows the schematic of the HPCVD system. The main process chamber is connected with the load-lock chamber on the bottom and an induction heater is attached on the outer periphery of the process chamber. A susceptor made from stainless steel is placed coaxially inside the reactor as shown in Fig. 2b. There is a gate-valve between the process chamber and the load-lock chamber. This valve protects the process chamber from contamination during the synthesis of MgB$_2$ films. HPCVD is capable of producing high quality MgB$_2$ films free of impurities, such as MgO and oxygen. In PIT-processed conductors, MgO is the main obstacle for the superconducting current transfer from one MgB$_2$ grain to adjacent grains and results in lower $J_c$ values [65,66].

The critical current density ($J_c$) of the order of $10^7$ A cm$^{-2}$ at 5 K in a zero magnetic field is obtained for HPCVD MgB$_2$ films [67,68]. Zhuang et al. [69] report the observation of a very high $J_c(0)$ of $1.6 \times 10^8$ A cm$^{-2}$ by the transport measurement in a clean-limit MgB$_2$ film grown by HPCVD, the highest self-field critical current density observed so far in MgB$_2$. These high $J_c$ values in HPCVD films suggest that a significant improvement in current carrying capability is possible in MgB$_2$. In addition, strong enhancement of critical current density [70] and record high values of upper critical field ($H_{c2}$) over 60 T [71] have been reported for carbon-doped MgB$_2$ films produced by HPCVD. Moreover, the deposition of MgB$_2$ films on the metallic substrates, such as stainless steel (SS), Nb and Cu have been demonstrated by Chen et al. using HPCVD [72]. In MgB$_2$ superconductor, it is well known that the grain boundaries are transparent to current flow [3]. In addition, grain boundaries in MgB$_2$ are proven to be the effective flux-pinning centers for enhancing $J_c$ [60]. This is very advantageous compared to cuprates high-$T_c$ superconductors where grain alignment is essential to improve grain coupling and to obtain high $J_c$ values.

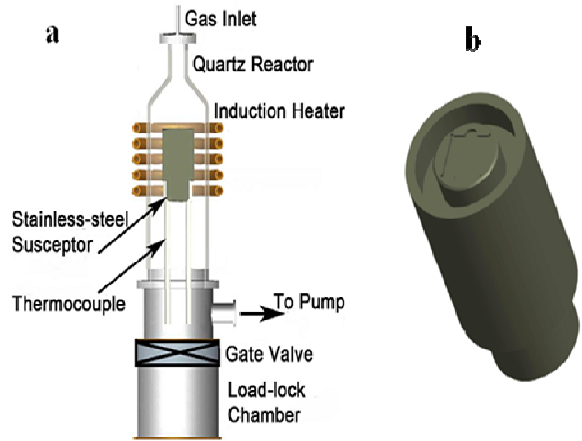

Fig. 2. (a) Schematic diagram of the HPCVD system. (b) The susceptor on which substrate is mounted on the top. About 6.5g of Mg can be accommodated around the substrate holder, good enough to maintain high Mg vapor pressure during deposition.

Therefore, the direct deposition of $MgB_2$ onto the substrate is possible and can avoid the necessity of buffer layers to reduce the cost as well as the complexity of the overall deposition process (Fig. 1b). The micron-thick polycrystalline $MgB_2$ films on stainless steel substrate and Nb sheets have been reported by Li et al. [73] and Zhuang et al. [74], respectively using the same method of HPCVD. The polycrystalline $MgB_2$ films on flexible YSZ substrates deposited by HPCVD has been reported by Pogrebnyakov et al. [75] with $T_c$ of 39 K and self-field $J_c$ over $10^7$ A cm$^{-2}$. The repeated bending over a radius of 10 mm does not change the superconducting properties of the films including $T_c$ and $J_c$. These good superconducting and mechanical properties are important for $MgB_2$ coated-conductor wires. Therefore, the impact of HPCVD technique along coated conductor approach is of great interest for the fabrication of $MgB_2$ tapes and wires.

## 4. Results on $MgB_2$/Cu (0 0 1) superconducting tapes

In order to fabricate long length $MgB_2$ tapes and wires with high critical current densities the selection of substrate material is very important [27,42]. In PIT process, the Fe sheath material most commonly used for the fabrication of $MgB_2$ tapes and wires has low thermal conductivity and is not suitable from the aspect of the high thermal stability of the tapes and wires. Superconducting $MgB_2$ wires and tapes have also been fabricated using Cu sheaths by PIT method [27,76–81]. Here we selected textured Cu (0 0 1) tape as a substrate due to its high thermal and electrical conductivity. Also Cu was found to be very helpful in the accelerated formation of the $MgB_2$ phase at low temperatures [82]. Although Cu reacts with Mg or $MgB_2$ [77] at high temperatures, in contrast to magnetic materials like Fe and Ni, this reaction can be reduced by lowering the growth temperature.

### 4.1. Fabrication of $MgB_2$/Cu tapes

In previous studies the growth of $MgB_2$ films on Cu substrates have been reported [83,84], whereas there is no report on the deposition of $MgB_2$ films on textured Cu (0 0 1) tapes. The textured Cu tapes have been prepared by one hour heat treatment of Cu ingot in a tube furnace at 900 ºC under a flow of a mixture of argon and hydrogen (5%) gases [85]. The thickness of as-prepared Cu tapes was ~90 μm and the tapes used in this study were cut into sizes of $10 \times 10$ mm$^2$.

The $MgB_2$ coated superconducting tapes were fabricated by depositing $MgB_2$ films on Cu tapes using the HPCVD process. In this process, the Cu (0 0 1) tape was placed on top surface of a susceptor and Mg solid pieces were placed around it. For the deposition of $MgB_2$ film on a Cu tape, the reactor was firstly evacuated to a base pressure of ~10$^{-3}$ Torr using rotary pump and purged several times by flowing high purity Ar and $H_2$ gases. Prior to the film growth, the susceptor along with Cu tape and Mg pieces were inductively heated towards the set temperature under a fixed reactor pressure in ambient $H_2$ gas. Upon reaching the set temperature, a boron precursor gas, $B_2H_6$ (5% in $H_2$) was introduced into the reactor to initiate the film growth. The flow rates were 50 sccm for the $H_2$ carrier gas and 5 sccm for the $B_2H_6/H_2$ mixture. Finally, the deposited $MgB_2$/Cu tape was cooled down to room temperature in a flowing $H_2$ carrier gas. The $MgB_2$/Cu tapes were fabricated in the temperature range of 460 to 520 ºC. The thickness of all fabricated $MgB_2$ tapes was ~1.3 μm with a growth rate of 65 nm min$^{-1}$.

### 4.2. XRD analysis

The X-ray diffraction (XRD) patterns of pure Cu (0 0 1) tape and $MgB_2$ superconducting tapes fabricated at various temperatures are shown in Fig. 3. Only the (0 0 $l$) peaks of $MgB_2$ are observed, indicating a $c$-axis-oriented crystal structure normal to the Cu (0 0 1) tape. The peaks of Cu tape are marked with asterisk (*). The $MgCu_2$ impurity phase is detected in all samples which come from the reaction between Mg and Cu tape during the film deposition process. It is noted that the amount of $MgCu_2$ impurity phase increases with increasing the growth temperature, implying that Cu is more reactive with Mg at higher temperatures.

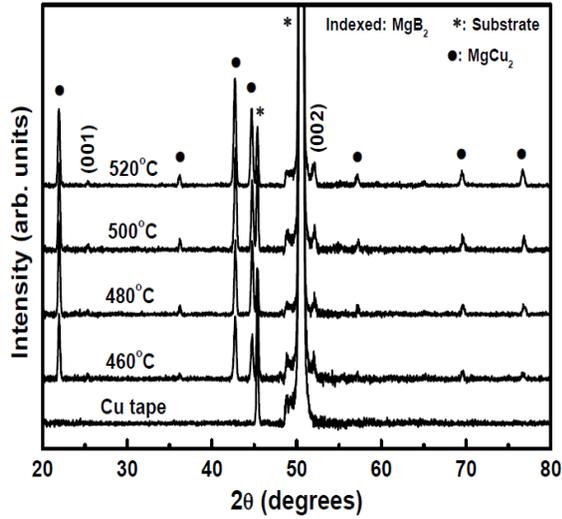

Fig. 3. X-ray diffraction patterns of textured Cu (0 0 1) tape and MgB$_2$ films grown on Cu (0 0 1) tapes at temperatures of 460–520 ºC.

*4.3. Temperature versus resistance curves*

The temperature dependences of the normalized resistances for MgB$_2$ films on textured Cu tapes are plotted in Fig. 4. The growth conditions, the superconducting transition temperature $T_c$, and $T_{c,zero}$ of MgB$_2$ superconducting tapes are summarized in Table 1. The tapes exhibited critical temperatures ranging between 36 and 38 K with superconducting transition width ($\Delta T_c$) of about 0.3–0.6 K. Among our samples, the tape fabricated at a temperature of 500 ºC has the highest $T_c$ of 37.5 K, which is comparable with the data reported by Li et al. [84] and much higher than that obtained for the MgB$_2$ films on Cu plates by Masuda et al. [23]. It is observed that the $T_c$ of the MgB$_2$ tapes decreased as we lowered the growth temperature. However, the tape prepared at a temperature of 520 ºC has lower $T_c$ than the other tapes but shows a sharp superconducting transition.

**Table 1**

The growth conditions, $T_c$ and $T_{c,zero}$ of MgB$_2$ coated superconducting tapes which are fabricated by growing MgB$_2$ films on textured Cu (0 0 1) tapes.

| $T_s$ (ºC) | $P$ (Torr) | $B_2H_6$ (sccm) | $H_2$ (sccm) | $T_c$ (K) | $T_{c,zero}$ (K) |
|---|---|---|---|---|---|
| 460 | 2 | 5 | 50 | 37.0 | 36.4 |
| 480 | 5 | 5 | 50 | 37.2 | 36.7 |
| 500 | 10 | 5 | 50 | 37.5 | 37.2 |
| 520 | 20 | 5 | 50 | 36.7 | 36.3 |

*4.4. SEM analysis*

Fig. 5 shows the surface morphologies of MgB$_2$ tapes observed by scanning electron microscopy (SEM). The MgB$_2$ grains are well connected for the tapes deposited at low temperatures of 460 and 480 ºC, as can be seen in Fig. 5a and b. The hexagonal crystallites of about 300–500 nm in size were observed for the tape grown at 460 ºC. However, the grain sizes of the tapes grown at high temperatures of 500 and 520 ºC are irregular and the connectivity between the MgB$_2$ grains is very poor. Fig. 5c and d, reflects the severe reactions of Mg vapor with Cu tape such as eutectic melting or diffusive movement between Mg vapor and Cu tape [86].

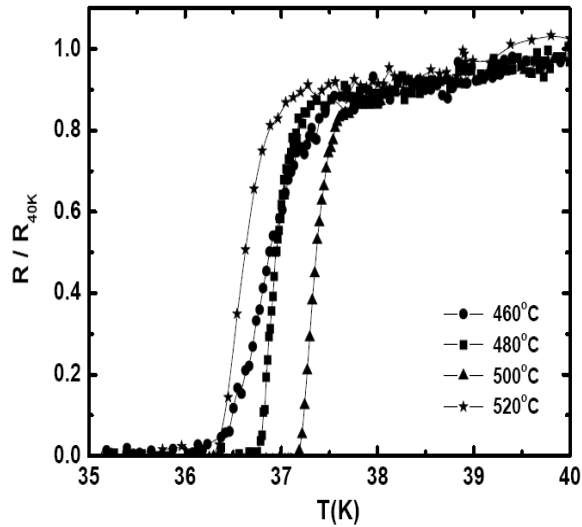

Fig. 4. Temperature dependences of the normalized resistances for MgB$_2$ coated superconducting tapes grown at various temperatures.

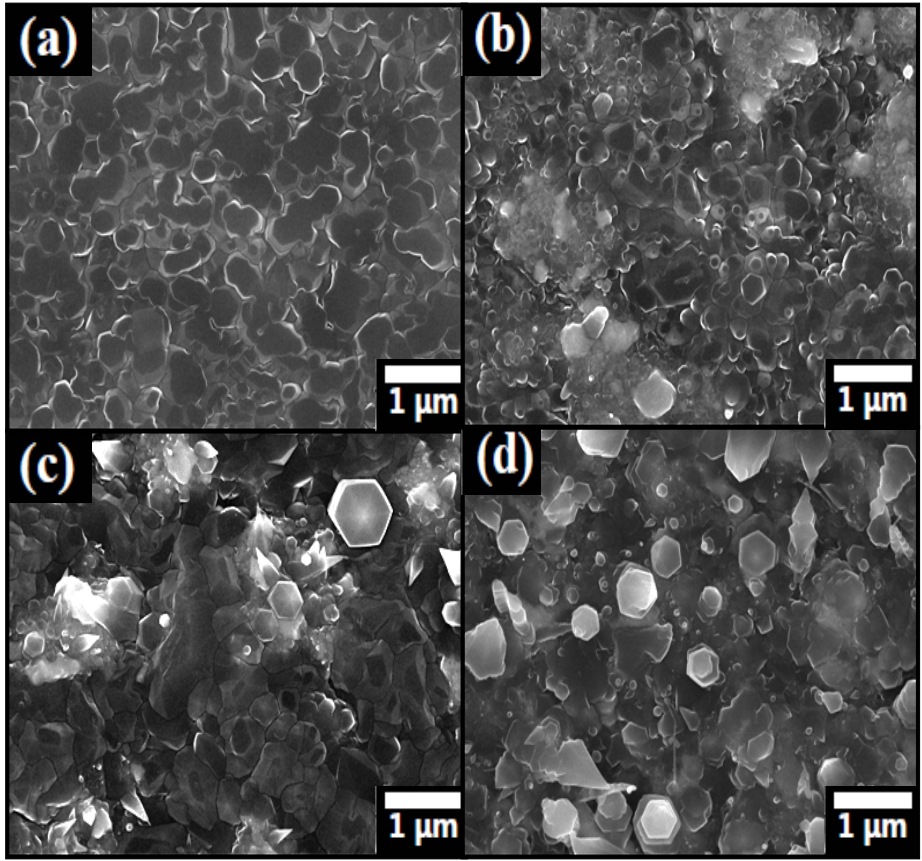

Fig. 5. SEM images of $MgB_2$/Cu tapes grown at various temperatures of (a) 460 °C, (b) 480 °C, (c) 500 °C, and (d) 520 °C.

## 4.5. Effect of growth temperature on $J_c$

The critical current density ($J_c$) is evaluated from the magnetic hysteresis (*M–H*) loops by using the Bean's critical state model. Fig. 6 shows the magnetic field dependence of the $J_c$ at 5 and 20 K for the $MgB_2$ superconducting tapes grown at various temperatures from 460 to 520 ºC. Among all the tapes, the highest $J_c$ of $1.34 \times 10^5$ A/cm$^2$ at 5 K under 3 T is obtained for the tape prepared at a temperature of 460 ºC. The $J_c$ of our tape over the wide field range is higher than the previous reported data for the film prepared on metallic Cu substrate by short-time sintering method [83] and the films deposited on polycrystalline Cu substrates by electron beam evaporation technique and HPCVD method [23,72]. Furthermore, on increasing the growth temperature a decrease in $J_c$ under applied field was observed for $MgB_2$ tapes. In 3 T field, the $J_c$ of the $MgB_2$ tape grown at 460 ºC is about 2.6 times higher than that grown at 520 ºC. This could be attributed to the high density of grain boundaries for the tape grown at 460 ºC, which may act as effective flux pinning centers. Another reason for the decrease of $J_c$ of 520 ºC sample is most likely due to increase of the extent of $MgCu_2$ impurity phase as we observed in XRD. The presence of these impurity phases reduces the density of superconducting $MgB_2$ and also acts as an obstacle against the flow of supercurrent [70,87]. Therefore, the tape grown at 460 ºC shows higher $J_c$ than that grown at 520 ºC.

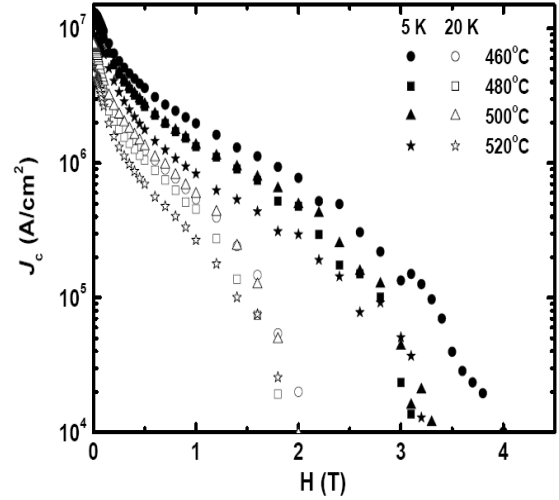

Fig. 6. Magnetic field dependence of critical current density ($J_c$) at 5 K (solid symbols) and 20 K (open symbols) for $MgB_2$ coated superconducting tapes fabricated at various temperatures.

*4.6. Flux pinning and SiC doping*

The self-field $J_c$ values as high as in the HPCVD $MgB_2$ films are obtained for the HPCVD $MgB_2$/Cu tapes. However, $J_c$ of pristine $MgB_2$/Cu tapes is decreased with an increasing external magnetic field because of their poor flux pinning properties [88]. To increase the potential use of $MgB_2$ superconductor in the practical applications, enhancement in the critical current density under high magnetic fields is essential. In these applications, strong pinning of vortices is an important requirement for an operation with low noise and lower power loss. Basically, the Lorentz force, $F_L = J \times B$, causes vortices to move in the superconductor, where $J$ is the applied current density and $B$ is the average magnetic induction inside the superconductor. If we understand the optimal pinning mechanism, we can obtain a higher $J_c$ because pinning sites, such as point defects, planar defects, etc., block the flux flow against the Lorentz force. The flux pinning force density ($F_p$) is defined as the product of the critical current density ($J_c$) and the induced magnetic field ($B$) in the sample, $F_p = -J_c \times B$. Pinning of vortices can be achieved by creating pinning sites, such as point defects [89], planar defects [90] and impurities [5,87].

Many researchers have attempted to improve the flux pinning behavior through several types of processes, such as proton [91] and neutron irradiation [92,93], oxygen alloying of $MgB_2$ thin films [16] and chemical doping using different metallic and nonmetallic phases, and nanoparticle admixing [94–96]. These studies have shown that chemical doping may be an effective and feasible approach for increasing critical current density of $MgB_2$ superconductors. Therefore, it is necessary to study the doping effect of suitable elements in $MgB_2$. Out of several dopants, carbon C [97] and carbon containing compounds, such as SiC [98], C and $CaCO_3$ co-addition [99] have been reported to be effective for improving the flux pinning properties of $MgB_2$. Employing HPCVD technique, so far carbon has been doped in $MgB_2$ films, using $(MeCp)_2Mg$ [100] and methane [101] as the carbon source. Recently, we report the doping of Cu and Ag in $MgB_2$ films by two step method [102]. Firstly, amorphous Cu and Ag-impurity layers were deposited on *c*-cut $Al_2O_3$ substrates at room temperature by pulsed laser deposition (PLD) system. Subsequently, $MgB_2$ films were grown on top of Cu or Ag/$Al_2O_3$ substrates at 560 and 600 ºC by using HPCVD. It was found that both Cu and Ag are effective for improving the flux pinning property due to creation of extra pinning sources in $MgB_2$ films. In this work, the amorphous SiC layer which is grown on metallic Cu (0 0 1) tape is used as an impurity for creating additional pinning sites in $MgB_2$ [103].

The PLD system was used for the deposition of SiC-impurity layers. The amorphous SiC-impurity layers of thickness 20 nm were deposited on the textured Cu tapes by PLD at room temperature with a background pressure of $\sim10^{-6}$ Torr. Laser beams were generated using a Lambda Physik KrF excimer laser ($\lambda$ = 248 nm). The pulse energy was set at 300 mJ with a repetition rate of 8 Hz. The SiC target (99.99% purity) was used for the deposition of impurity layers. The PLD system used in this study is described in more detail [104,105]. After depositing the SiC-impurity layers on the Cu tapes, we used HPCVD system for the fabrication of $MgB_2$ coated superconducting tapes by growing $MgB_2$ films on SiC/Cu tapes.

In this process, the Cu tape coated with SiC-impurity layer was placed on top surface of a susceptor and Mg chips were placed around it. For the deposition of $MgB_2$ film on a SiC/Cu tape, the reactor was firstly evacuated to a base pressure of $\sim10^{-3}$ Torr using rotary pump and purged several times by flowing high purity argon and hydrogen gases. Prior to the film growth, the susceptor along with SiC/Cu tape and Mg chips were inductively heated towards the set temperature under a reactor pressure of 50 Torr in ambient $H_2$ gas. Upon reaching the set temperature, a boron precursor gas, $B_2H_6$ (5% in $H_2$) was introduced into the reactor to initiate the film growth. The flow rates were 90 sccm for the $H_2$ carrier gas and 10 sccm for the $B_2H_6/H_2$ mixture. Finally, the fabricated $MgB_2$/SiC/Cu tape was cooled down to room temperature in a flowing $H_2$ carrier gas. Since the critical properties, such as $T_c$, $J_c$, and $H_{c2}$ of superconductors are strongly influenced by heat treatment conditions. Therefore, the performance of any superconductor can be improved further by optimizing the growth temperature. Thus, it is necessary to investigate the heat treatment effect on the $J_c$–$H$ behavior of $MgB_2$/SiC/Cu tapes. The $MgB_2$/SiC/Cu tapes were fabricated in the temperature range of 460 to 600 °C. A pure $MgB_2$/Cu tape was also prepared at a temperature of 460 °C for comparison. The thickness of all fabricated $MgB_2$ tapes was ~2 μm with a growth rate of 0.2 μm min$^{-1}$.

The $J_c$ ($H$) curves for $MgB_2$/Cu and $MgB_2$/SiC/Cu tapes as a function of growth temperature measured at 5 and 20 K are shown in Fig. 7a and b, respectively. We can see that both at 5 and 20 K, the self-field $J_c$ for SiC-coated $MgB_2$ tapes is higher than that of pure $MgB_2$ tape. It is well established that SiC doping in $MgB_2$ using the conventional in situ PIT technique shows a decrease in $T_c$ and reduction of the self-field $J_c$ due to the high level of C substitution on the B sites [98,106]. Recently Li et al. obtained enhancement in the self-field $J_c$ for SiC-$MgB_2$ composite caused by thermal strain on the interface between SiC and $MgB_2$ during the diffusion process, when no chemical reaction between SiC and $MgB_2$ was observed [107]. No reduction of the $T_c$ and the absence of any chemical reaction between SiC-impurity layer and $MgB_2$ might be the reason for improvement in the self-field $J_c$ of our $MgB_2$/SiC/Cu tapes. Furthermore, on increasing the deposition temperature an increase in $J_c$ under applied field was observed for SiC-coated tapes. The optimal growth temperature for the strong improvement in $J_c$ was observed to be 540 °C. These results are in contrast to our previously reported data for $MgB_2$/Cu tapes, where $MgB_2$ films were directly deposited on the textured Cu tapes and they showed a decrease in $J_c$ with increasing growth temperature [88]. The enhanced $J_c$ for $MgB_2$/SiC/Cu tapes could be due to the improved flux pinning by grain boundaries along with the additional defects created by SiC-impurity layer which act as effective flux pinning centers [108]. It is also noted that the in-field $J_c$ of $MgB_2$/Cu tape around 4 and 2 T at 5 and 20 K, respectively is comparable with the $MgB_2$/SiC/Cu tapes. This might be due to the smaller grain size of 350 nm for $MgB_2$/Cu tape as a result this sample has a high density of grain boundaries, and grain boundaries are known to be the main pinning source in $MgB_2$ superconductor [109].

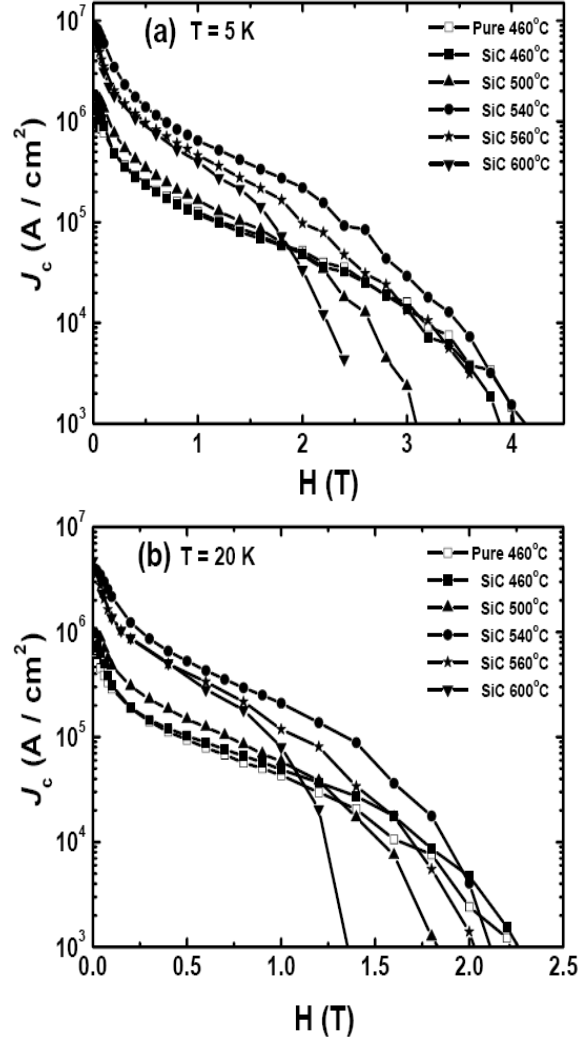

Fig. 7. The $J_c$ ($H$) curves as a function of growth temperature for $MgB_2$/Cu and $MgB_2$/SiC/Cu tapes measured at (a) 5 K, and (b) 20 K. The optimal growth temperature for the strong improvement in $J_c$ for $MgB_2$/SiC/Cu tapes was observed to be 540 °C.

Fig. 8 shows the magnetic field dependences of the flux pinning force densities ($F_p$) at 5 K for $MgB_2$/Cu and $MgB_2$/SiC/Cu tapes grown at various temperatures, where $F_p$ is derived from the corresponding $J_c(H)$ data (Fig. 7a). The $F_p$–$H$ graph clearly shows an increase in the flux pinning force density with increasing growth temperature for SiC-coated $MgB_2$ tapes as compared to $MgB_2$/Cu tape. The largest enhancement of $F_p$ is observed for the $MgB_2$/SiC/Cu tape fabricated at 540 °C which has a maximum $F_p$ value, about 5.4 times higher than the $MgB_2$/Cu tape. It indicates that the $MgB_2$/SiC/Cu tapes have higher density of pinning centers, such as grain boundaries and additional defects as a consequence of SiC-impurity layer as compared to $MgB_2$/Cu tape.

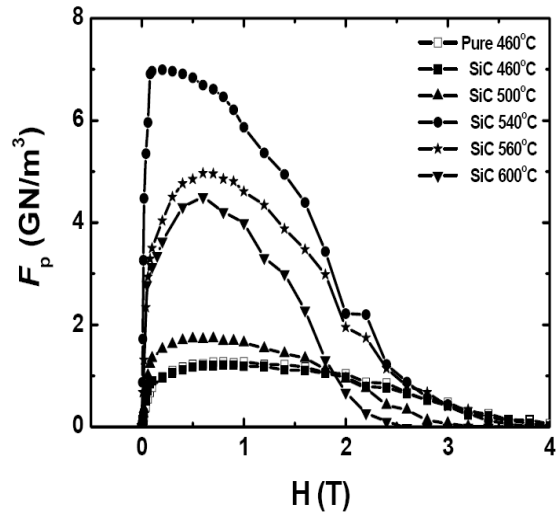

Fig. 8. Magnetic field dependence of flux pinning force density ($F_p$) at 5 K for $MgB_2$/Cu and $MgB_2$/SiC/Cu tapes grown at various temperatures.

## 5. Results on thick $MgB_2$/Hastelloy superconducting tapes

The role of substrate material is very crucial in the fabrication of $MgB_2$ conductors to obtain high $J_c$ values. The $J_c$ of our $MgB_2$/Cu tapes decreased with an increasing applied field because of their poor flux pinning as well as high reactivity of Mg with Cu tape, as a result, reduction of the superconducting cross sectional area and hence the decrease of $J_c$ values [88]. Thus, we need to use a metallic substrate which does not react with Mg and B. Komori et al. obtained very high in-field transport critical current density for $MgB_2$ grown on YSZ buffered Hastelloy tapes without any chemical reactions between Mg and Hastelloy tapes [18]. In addition, Hastelloy tape has good material properties, such as good flexibility, good conductivity, and lower corrosiveness. Thus, Hastelloy tape could be expected as a good metallic substrate for $MgB_2$ tape conductor fabrications.

For coated conductor applications, it is an important to have a thick superconducting film at high growth rate [48, 110], so that the high critical current density ($J_c$) translates to high engineering critical current density ($J_e$) of a conductor [85]. To see the effect of film thickness and deposition rate on the superconducting properties, mainly on $J_c$ of $MgB_2$ superconductor a study is needed to be performed. Wang et al. [111] and Hanna et al. [112] have reported the correlation between film thickness and critical current density for the $MgB_2$ films fabricated on single-crystal MgO (111) and (0001) 6H-SiC substrates, respectively by *ex-situ* methods. In order to use $MgB_2$ for coated-conductor applications, a clear idea about the relationship between film thickness and critical current density for the $MgB_2$ films fabricated on metallic substrates is necessary. Recently, we have investigated the influence of $B_2H_6$ gas mixture flow rate and deposition time on superconducting properties of $MgB_2$ films deposited on metallic Hastelloy tapes by HPCVD [113].

### 5.1. Fabrication of thick $MgB_2$/Hastelloy tapes

In this work, we use the polycrystalline Hastelloy tapes of thickness ~45 μm and the tapes were cut into sizes of 10 × 10 mm$^2$. The $MgB_2$ coated superconducting tapes were prepared by growing $MgB_2$ films on Hastelloy tapes with different $B_2H_6$ gas mixture flow rates and for different deposition times. All tapes were fabricated at a temperature of 520 °C under

## Table 2

The growth conditions, thickness, $T_c$ and $T_{c,zero}$ of MgB$_2$/Hastelloy tapes fabricated at a temperature of 520 °C under a pressure of 200 Torr.

| H$_2$ (sccm) | B$_2$H$_6$ (sccm) | Deposition time (min.) | Thickness (μm) | $T_c$ (K) | $T_{c,zero}$ (K) |
|---|---|---|---|---|---|
| 90 | 10 | 10 | 0.5 | 38.5 | 38.1 |
| 80 | 20 | 10 | 1.5 | 38.8 | 38.2 |
| 70 | 30 | 10 | 2.0 | 38.6 | 38.2 |
| 90 | 10 | 30 | 1.6 | 38.9 | 38.6 |
| 80 | 20 | 30 | 4.5 | 39.6 | 39.1 |
| 70 | 30 | 30 | 6.0 | 39.3 | 39.0 |

a pressure of 200 Torr. The B$_2$H$_6$ gas flow rate was varied from 10 to 30 sccm at H$_2$ of 90–70 sccm, respectively under different deposition times as summarized in Table 2. For the deposition of MgB$_2$ film on Hastelloy tape, the reactor was firstly evacuated to a base pressure of $\sim 10^{-3}$ Torr using rotary pump and purged several times by flowing high purity Ar and H$_2$ gases. Prior to the film growth, the susceptor along with Hastelloy tape and Mg chips were inductively heated towards the set temperature in ambient H$_2$ gas. Upon reaching the set temperature, a hydrogen-diluted B$_2$H$_6$ (5% in H$_2$) gas was introduced into the reactor to initiate the film growth. Finally, the fabricated MgB$_2$/Hastelloy tape was cooled down to room temperature in a flowing H$_2$ carrier gas. The MgB$_2$ layer thicknesses range between 0.5 and 6.0 μm.

### 5.2. XRD analysis

Fig. 9 shows the X-ray diffraction patterns for pure Hastelloy tape and MgB$_2$/Hastelloy tapes fabricated at 520 °C under various B$_2$H$_6$ gas mixture flow rates and deposition times. The MgB$_2$ tapes are polycrystalline in nature, as expected because of the use of polycrystalline Hastelloy tapes. It should be noted that the MgB$_2$ tapes deposited for longer time have preferred orientation along the (0 0 l) plane in addition to diffraction peaks of different MgB$_2$ planes compared to those deposited for shorter time. The c-axis lattice parameter was calculated using the (0 0 2) reflections of the tapes. For the tapes deposited under various B$_2$H$_6$ gas mixture flow rates and deposition times with the thicknesses of 0.5, 1.6, 2.0, and 6.0 μm (Table 2), it is observed to be 3.526, 3.525, 3.524, and 3.521 Å, respectively. The compression of c-axis lattice parameter with the increase of MgB$_2$ layer thickness indicates that these tapes are under tensile strain [114]. Moreover, there is no indication for any chemical reaction between MgB$_2$ and Hastelloy tape, neither any impurity or secondary phases, such as Mg, MgO, and MgB$_4$ were observed.

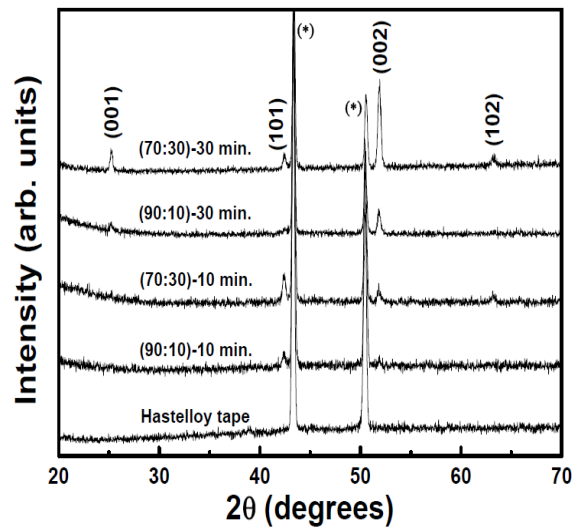

Fig. 9. X-ray diffraction patterns of pure Hastelloy tape and MgB$_2$/Hastelloy tapes grown at a temperature of 520 °C under various growth conditions. The Hastelloy tape peaks are marked with asterisk (*).

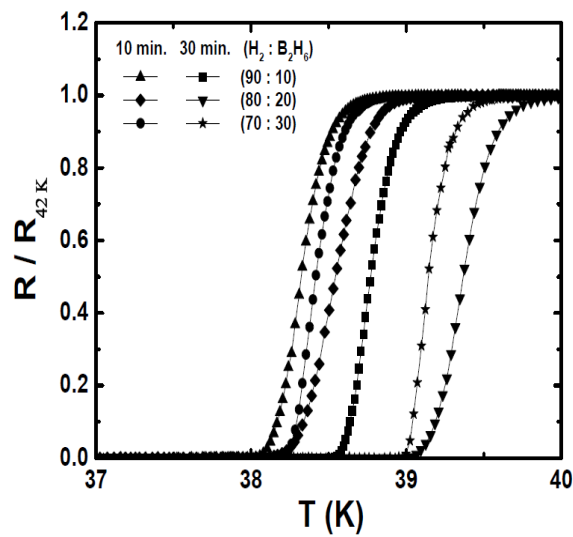

Fig. 10. Temperature dependences of the normalized resistances for MgB$_2$/Hastelloy tapes fabricated at various B$_2$H$_6$ gas mixture flow rates and deposition times.

*5.3. Temperature versus resistance curves*

The temperature dependences of the normalized resistances for the MgB$_2$/Hastelloy tapes are plotted in Fig. 10. The MgB$_2$ tapes have critical temperature ranging between 38.5 and 39.6 K with superconducting transition width ($\Delta T_c$) of 0.3–0.6 K. From Fig. 10, we can see that there is a little effect of B$_2$H$_6$ gas mixture flow rate on $T_c$, and the MgB$_2$ tapes deposited for longer time have higher $T_c$ than those deposited for shorter time. It indicates that $T_c$ is depending on the thickness of MgB$_2$ layer. The thicker films show higher $T_c$ than the thinner films have been reported for the films prepared by the same method of HPCVD [115]. The $T_c$ of our tapes is higher by 2 K than we previously obtained for MgB$_2$/Cu tapes [88], and it is much higher than that reported by Komori et al. of 29 K for post annealed MgB$_2$/YSZ/Hastelloy tapes [18]. The high $T_c$ of our samples is most probably due to high purity of the samples and the tensile strain in the films. The tensile strain is reported to be the possible cause of higher $T_c$ in MgB$_2$ films, and the thicker the films the larger the tensile strain [114].

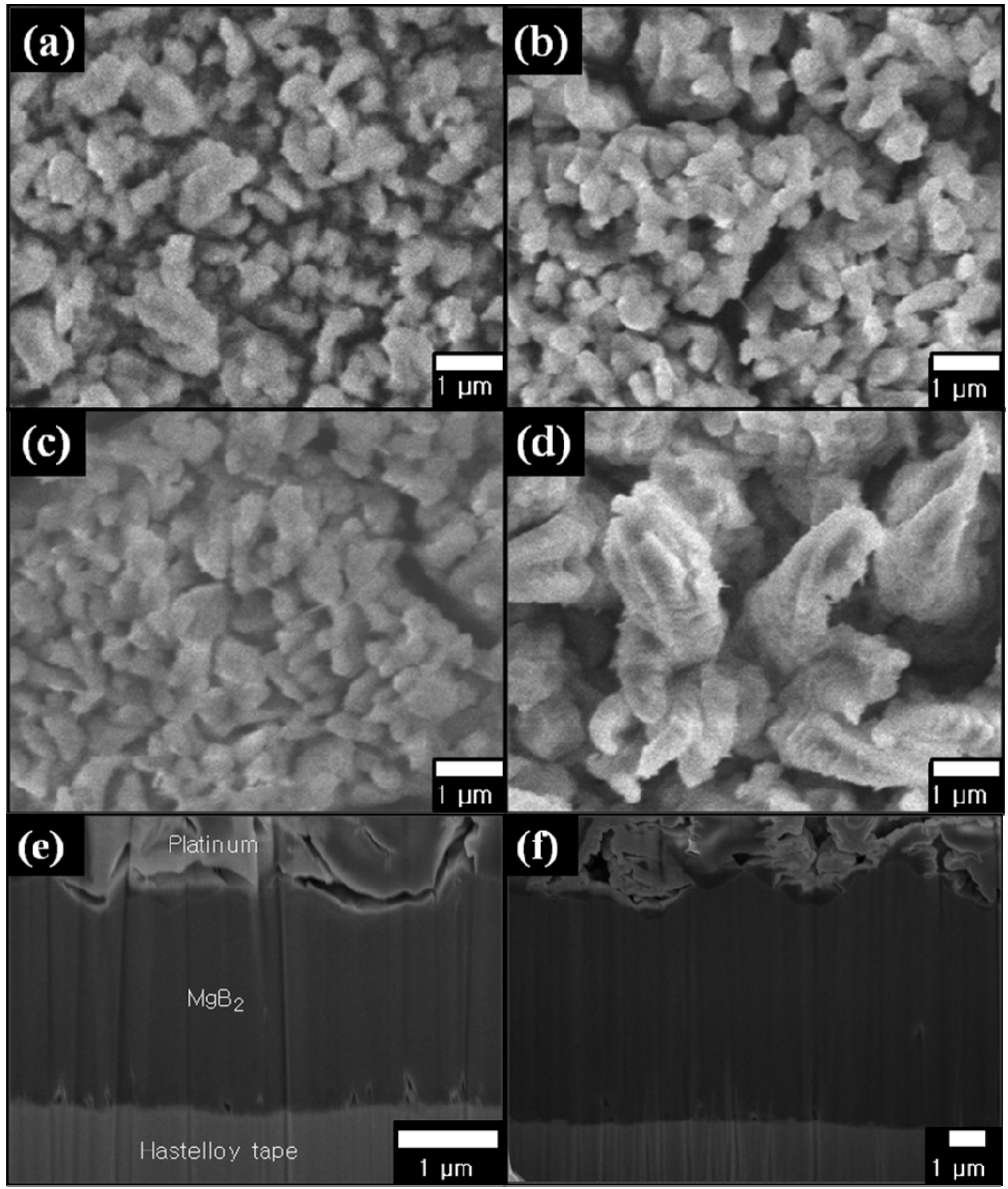

Fig. 11. The surface morphologies of MgB$_2$/Hastelloy tapes deposited at different H$_2$:B$_2$H$_6$ gas mixture flow rate and deposition time of (a) 90:10-10 min., (b) 70:30-10 min., (c) 90:10-30 min., and (d) 70:30-30 min. The cross-sectional views of the MgB$_2$ tapes deposited at (e) 70:30-10 min. and (f) 70:30-30 min.

*5.4. Microstructure observations*

The plane views and the cross-sectional views of the MgB$_2$/Hastelloy tapes were observed by scanning electron microscopy (SEM) and focused ion beam (FIB) technique, respectively. The surface morphologies and the cross-sectional views of MgB$_2$/Hastelloy tapes fabricated with different B$_2$H$_6$ gas mixture flow rates and for different deposition times are shown in Fig. 11a–f. Grains of various orientations could be seen in the images, suggesting that the MgB$_2$ tapes are polycrystalline, which is consistent with XRD results. In addition, the surfaces of the MgB$_2$ tapes look porous. The tape deposited with high B$_2$H$_6$ flow rate and for longer time, Fig. 11d, results in the tilted and isolated grain growth of few μm in sizes. It indicates that the microstructure degradation occurs for the tapes deposited with high B$_2$H$_6$ flow rate and for longer time. Fig. 11e and f are the cross-sectional views of MgB$_2$/Hastelloy tapes. Porosity was observed at the interfaces of Hastelloy tapes and MgB$_2$ layers whereas the inner areas of the MgB$_2$ layers are composed of dense structure without porosities.

*5.5. MgB$_2$ layer thickness dependence of $J_c$*

Fig. 12 shows the magnetic field dependence of $J_c$ at 5 and 20 K for MgB$_2$ tapes fabricated with various B$_2$H$_6$ flow rates and deposition times. In order to explain the $J_c(H)$ behavior, we use the MgB$_2$ tape thickness of 0.5, 1.5, 2.0, and 6 μm, as shown in Table 2. On increasing the thickness an increase in $J_c$ was observed for MgB$_2$ tapes. The optimal thickness for the strong improvement in $J_c$ was observed to be 2.0 μm. These results are in contrast to the data reported by Wang et al. [111] and Hanna et al. [112], where they obtained a decrease in $J_c$ with increasing MgB$_2$ film thickness, in the similar trend as observed in YBCO-coated conductors [116]. In our previous study, we report that the thick MgB$_2$ film has higher $J_c$ than the thin one, due to much higher density of pinning centers with weak intercolumnar regions in the thicker film [57]. However, further increasing the MgB$_2$ layer thickness over 2 μm, caused a reduction of $J_c$, the rapid fall of $J_c$ can be seen for the 6-μm-thick MgB$_2$ tape. The decrease of $J_c$ is most likely due to the degradation of the microstructure and the growth of larger grain sizes, as a result, less density of pinning centers. Our MgB$_2$/Hastelloy tapes showed very high $J_c$ values over the wide field range than the previously reported data on other metallic substrates, such as Cu, SS, and Nb [22,23,72,88].

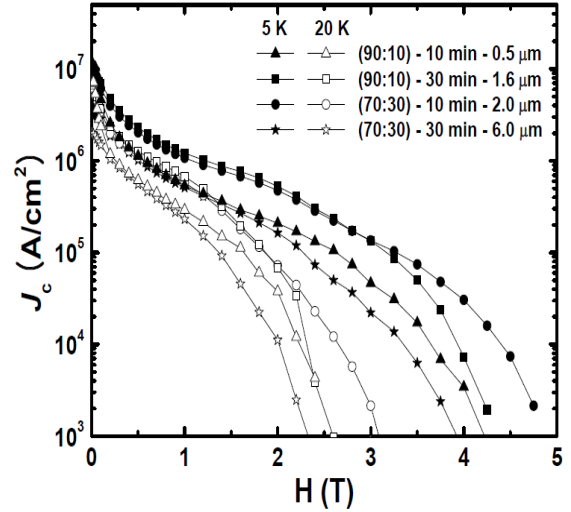

Fig. 12. The critical current density as a function of magnetic field measured at 5 K (solid symbols) and 20 K (open symbols) for MgB$_2$/Hastelloy tapes grown at different B$_2$H$_6$ gas mixture flow rates and deposition times.

## 6. Conclusions

In this article, we report our recent results on MgB$_2$ coated superconducting tapes fabricated using coated conductor approach by HPCVD technique. Our HPCVD MgB$_2$ tapes showed self-field $J_c$ values at 5 K of the order of $10^7$ A/cm$^2$ as high as already obtained in the MgB$_2$ films and 1–2 order higher than those reported for PIT-processed MgB$_2$ conductors. The MgB$_2$/Cu tape fabricated at 460 ºC showed the $J_c$ values of $J_c$(20 K, 0 T) ~6.5 × 10$^6$ A/cm$^2$ and $J_c$(20 K, 1.5 T) ~1.8 × 10$^5$ A/cm$^2$. The SiC-doped MgB$_2$/Cu tapes showed very high $J_c$ values as compared to MgB$_2$/Cu tapes and exhibited opposite trend in the dependence of $J_c$ with growth temperature. The strong improvement in $J_c$ for the MgB$_2$/SiC/Cu tapes could be attributed to the improved flux pinning by the additional defects created by SiC-impurity layer along with the MgB$_2$ grain boundaries. The MgB$_2$/Hastelloy tapes grown at 520 ºC with optimal thickness of 2 μm showed the highest $J_c$ values of $J_c$(20 K, 0 T) ~5.8 × 10$^6$ A/cm$^2$ and $J_c$(20 K, 1.5 T) ~2.4 × 10$^5$ A/cm$^2$. Even the MgB$_2$/Hastelloy tape as thick as 6 μm showed high $J_c$ values of $J_c$(20 K, 0 T) ~1.9 × 10$^6$ A/cm$^2$ and $J_c$(20 K, 1 T) ~2.3 × 10$^5$ A/cm$^2$. The HPCVD MgB$_2$ tapes on Hastelloy showed very high $J_c$ values over the wide field range than for those deposited on other metallic substrates, such as Cu, SS, and Nb. In addition, no chemical reaction was observed between Mg and Hastelloy tape. Thus, the Hastelloy tape could be a promising metallic substrate for the fabrication of MgB$_2$ tape conductors for practical applications. These results on our HPCVD MgB$_2$ coated superconducting tapes suggest that practical applications of MgB$_2$ at 20 K in low magnetic fields are promising. Although the self-field and low field $J_c$ values are high in these MgB$_2$ tapes than the existing superconducting wires of NbTi and Nb$_3$Sn, further improvement in the current carrying capability under the applied field is essential to realize the potential of MgB$_2$ superconductors as high-field magnets. Therefore, the main challenge now is to transfer the already achieved high $H_{c2}$ and $J_c$ values in carbon-doped MgB$_2$ films to the conductor forms which are more suitable for large scale applications.

## Acknowledgements

This work was supported by Mid-career Researcher Program through National Research Foundation of Korea (NRF) grant funded by the Ministry of Education, Science & Technology (MEST) (No. 2010-0029136).

## References

[1] J. Nagamatsu, N. Nakagawa, T. Muranaka, Y. Zenitani, J. Akimitsu, Nature (London) 410 (2001) 63.
[2] W. N. Kang, H. J. Kim, E. M. Choi, C. U. Jung, S. I. Lee, Science 292 (2001) 1521.
[3] D. C. Larbalestier et al., Nature 410, (2001) 186.
[4] W. N. Kang, H. J. Kim, E. M. Choi, H. J. Kim, K. H. P. Kim, H. S. Lee, S. I. Lee, Phys. Rev. B 65 (2002) 134508.
[5] D. Larbalestier, A. Gurevich, D. M. Feldmann, A. Polyanskii, Nature (London) 414 (2001) 368.
[6] C. Buzea, T. Yamashita, Supercond. Sci. Technol. 14 (2001) R115.
[7] K. Vinod, R. G. A. Kumar, and U. Syamaprasad, Supercond. Sci. Technol. 20, R1 (2007).
[8] R. Flukiger , H. L. Suo, N. Musolino, C. Beneduce, P. Toulemonde, P. Lezza, Physica C 385 (2003) 286.
[9] M. J. Holcomb, Physica C 423 (2005) 103.
[10] H. Kumakura, A. Matsumoto, T. Nakane, and H. Kitaguchi, Physica C 456 (2007) 196.
[11] M. Tomsic, M. Rindfleisch, J. Yue, K. McFadden, D. Doll, J. Phillips, M. D. Sumption, M. Bhatia, S. Bohnenstiehl, E. W. Collings, Physica C 456 (2007) 203.
[12] V. Braccini, D. Nardelli, R. Penco, G.Grasso, Physica C 456 (2007) 209.
[13] J. M. Rowell, Supercond. Sci. Technol. 16 (2003) R17.
[14] A. Matsumoto, H. Kumakura, H. Kitaguchi, B. J. Senkowicz, M. C. Jewell, E. E. Hellstrom, Y. Zhu, P. M. Voyles, D. C. Larbalestier, Appl. Phys. Lett. 89 (2006) 132508.
[15] H. J. Kim, W. N. Kang, E. M. Choi, M. S. Kim, Kijoon H. P Kim, S. I. Lee, Phys. Rev. Lett. 87 (2001) 087002.
[16] C. B. Eom et al., Nature 411 (2001) 558.
[17] M. Paranthaman et al., Appl. Phys. Lett. 78 (2001) 3669.
[18] K. Komori, K. Kawagishi, Y. Takano, H. Fujii, S. Arisawa, H. Kumakura, M. Fukutomi, K. Togano, Appl. Phys. Lett. 81 (2002) 1047.
[19] J. D. Xu, S. F. Wang, Y. B. Zhou, Y. L. Zhou, Z. H. Chen, D. F. Cui, H. B. Lu, M. He, S. Y. Dai, G. Z. Yang, Supercond. Sci. Technol. 15 (2002) 1190.
[20] B. A. Glowacki, M. Majoros, M. Vickers, M. Eisterer, S. Toenies,H. W. Weber, M. Fukutomi, K. Komori, K. Togano, Supercond. Sci. Technol. 16 (2003) 297.
[21] V. Ferrando et al., Appl. Phys. Lett. 87 (2005) 252509.
[22] H. Abe, K. Nishida, M. Imai, H. Kitazawa, K. Yoshii, Appl. Phys. Lett. 85 (2004) 6197.
[23] K. Masuda, T. Doi, K. Fukuyama, S. Hamada, Y. Hakuraku, and H. Kitaguchi, IEEE Trans. Appl. Supercond. 17 (2007) 2895.
[24] P. C. Canfield, D. K. Finnemore, S. L. Budko, J. E. Ostenson, G. Lapertot, C. E. Cunningham, C. Petrovic, Phys. Rev. Lett. 86 (2001) 2423.
[25] J. D. DeFouw, D. C. Dunand, Appl. Phys. Lett. 83 (2003) 120.

[26] K. Togano, T. Nakane, H. Fujii, H. Takeya, H. Kumakura, Supercond. Sci. Technol. 19 (2006) L17.
[27] G. Grasso, A. Malagoli, C. Ferdeghini, S. Roncallo, V. Braccini, A. S. Siri, Appl. Phys. Lett. 79 (2001) 230.
[28] H. Kumakura, A. Matsumoto, H. Fujii, K. Togano, Appl. Phys. Lett. 79 (2001) 2435.
[29] C. H. Jiang, T. Nakane, H. Kumakura, Appl. Phys. Lett. 87 (2005) 252505.
[30] O. Perner, J. Eckert, W. Häßler, C. Fischer, J. Acker, T. Gemming, G. Fuchs, B. Holzapfel, L. Schultz, J. Appl. Phys. 97 (2005) 056105.
[31] S. K. Chen, K. A. Yates, M. G. Blamire, J. L. MacManus-Driscoll, Supercond. Sci. Technol. 18 (2005) 1473.
[32] X. Xu, M. J. Qin K. Konstantinov, D. I. Santos, W. K. Yeoh, J. H. Kim, S. X. Dou, Supercond. Sci. Technol. 19 (2006) 466.
[33] J. Jiang, B. J. Senkowicz, D. C. Larbalestier, E. E. Hellstrom, Supercond. Sci. Technol. 19 (2006) L33.
[34] H. L. Suo, C. Beneduce, M. Dhall, N. Musolino, J. Y. Genoud, R. Flukiger, Appl. Phys. Lett. 79 (2001) 3116.
[35] H. Yamadaa, M. Hirakawa, H. Kumakura, A. Matsumoto, H. Kitaguchi, Appl. Phys. Lett. 84 (2004) 1728.
[36] M. J. Qin, S. Keshavarzi, S. Soltanian, X. L. Wang, H. K. Liu, S. X. Dou, Phys. Rev. B 69 (2004) 012507.
[37] C. F. Liu, G. Yan, S. J. Du, W. Xi, Y. Feng, P. X. Zhang, X. Z. Wu, L. Zhou, Physica C 386 (2003) 603.
[38] H. Fang, Y. Y. Xue, Y. X. Zhou, A. Baikalov, K. Salama, Supercond. Sci. Technol. 17 (2004) L27.
[39] A. Serquis, L. Civale, D. L. Hammon, J. Y. Coulter, X. Z. Liao, Y. T. Zhu, D. E. Peterson, F. M. Mueller, Appl. Phys. Lett. 82 (2003) 1754.
[40] A. Serquis, L. Civale, D. L. Hammon, X. Z. Liao, J. Y. Coulter, Y. T. Zhu, D. E. Peterson, F. M. Mueller, J. Appl. Phys. 94 (2003) 4024.
[41] S. K. Chen, K. S. Tan, B. A. Glowacki, W. K. Yeoh, S. Soltanian, J. Horvat, S. X. Dou, Appl. Phys. Lett. 87 (2005) 182504.
[42] M. Kiuchi, K. Yamauchi, T. Kurokawa, E. S. Otabe, T. Matsushita, M. Okada, K. Tanaka, H. Kumakura, H. Kitaguchi, Physica C 412 (2004) 1189.
[43] T. Nakane, H. Kitaguchi, H. Kumakura, Appl. Phys. Lett. 88 (2006) 022513.
[44] W. K. Yeoh, S. X. Dou, Physica C 456 (2007) 170.
[45] M. A. Susner, M. D. Sumption, M. Bhatia, X. Peng, M. J. Tomsic, M. A. Rindfleisch, E. W. Collings, Physica C 456 (2007) 180.
[46] Y. Iijima, N. Tanabe, O. Kohno, and Y. Ikeno, Appl. Phys. Lett. 60, 769 (1992).
[47] D. Dimos, P. Chaudhari, and J. Mannhart, Phys. Rev. B 41, 4038 (1990).
[48] S. R. Foltyn, L. Civale, J. L. MacManus-Driscoll, Q. X. Jia, B. Maiorov, H. Wang, M. Maley, Nat. Mater. 6 (2007) 631.
[49] Y. Iwasa, D. C. Larbalestier, M. Okada R. Penco, M. D. Sumption, X. Xi, IEEE Trans. Appl. Supercond. 16 (2006) 1457.
[50] K. Ueda, M. Naito, Appl. Phys. Lett. 79 (2001) 2046.
[51] A. J. M. Van Erven, T. H. Kim, M. Muenzenberg, J. S. Moodera, Appl. Phys. Lett. 81 (2002) 4982.
[52] T. H. Kim, J. Kor. Phys. Soc. 49, (2006) L1881.
[53] H. Y. Zhai, H. M. Christen, L. Zhang, C. Cantoni, M. Paranthaman, B. C. Sales, D. K. Christen, D. H. Lowndes, Appl. Phys. Lett. 79 (2001) 2603.
[54] A. Brinkman et al., Physica C 353 (2001) 1.
[55] D. Mijatovic, A. Brinkman, G. Rijnders, H. Hilgenkamp, H. Rogalla, D. H. A. Blank, Physica C 372 (2002) 1258.
[56] X. H. Zeng, A. V. Pogrebnyakov, A. Kotcharov, J. E. Jones. X. X. Xi, E. M. Lysczek, J. M. Redwing, S. Y. Xu, Q. Li, J. Lettieri, D. G. Schlom W. Tian, X. Pan, Z. K. Liu, Nat. Mater. 1 (2002) 35.
[57] W. K. Seong, J. Y. Huh, W. N. Kang, J.-W. Kim, Y.-S. Kwon, N.-K. Yang, J.-G. Park, Chem. Vapor Depos. 13 (2007) 680.
[58] X. X. Xi et al., Physica C 456 (2007) 22.
[59] M. Okuzono, T. Doi, Y. Ishizaki, Y. Kobayashi, Y. Hakuraku, H. Kitaguchi, IEEE Trans. Appl. Supercond. 15 (2005) 3253.
[60] H. Kitaguchi, A. Matsumoto, H. Kumakura, T. Doi, H. Yamamoto, K. Saitoh, H. Sosiati, S. Hata, Appl. Phys. Lett. 85 (2004) 2842.
[61] M. Haruta, T. Fujiyoshi, T. Sueyoshi, K. Miyahara, T. Doi, H. Kitaguchi, S. Awaji, K. Watanabe, Supercond. Sci. Technol. 18 (2005) 1460.
[62] B. H. Moeckly, W. S. Ruby, Supercond. Sci. Technol. 19 (2006) L21.
[63] E. Monticone, C. Gandini, C. Portesi, M. Rajteri, S. Bodoardo, N. Penazzi, V. Dellarocca, R. S. Gonnelli, Supercond. Sci. Technol. 17 (2004) 649.
[64] J. Rowell, Nat. Mater. 1 (2002) 5.
[65] H. Fujii, H. Kumakura, K. Togano, Physica C 363 (2001) 237.
[66] S. Hata, T. Yoshidome, H. Sosiati, Y. Tomokiyo, N. Kuwano, A. Matsumoto, H. Kitaguchi, H. Kumakura, Supercond. Sci. Technol. 19 (2006) 161.
[67] X. H. Zeng et al., Appl. Phys. Lett. 82 (2003) 2097.
[68] W. K. Seong, W. N. Kang, Physica C 468 (2008) 1884.
[69] C. G. Zhuang, S. Meng, C. Y. Zhang, Q. R. Feng, Z. Z. Gan, H. Yang, Y. Jia, H. H. Wen, X. X. Xi, J. Appl. Phys. 104 (2008) 013924.
[70] J. Chen et al., Phys. Rev. B 74 (2006) 174511.
[71] V. Braccini et al., Phys. Rev. B 71 (2005) 012504.
[72] L-P Chen et al., Chin. Phys. Lett. 24 (2007) 2074.
[73] F. Li, T. Guo, K. Zhang, C. Chen, Q. Feng, Physica C 452 (2007) 6.
[74] C. Zhuang, D. Yao, F. Li, K. Zhang, Q. Feng, Z. Gan, Supercond. Sci. Technol. 20 (2007) 287.
[75] A. V. Pogrebnyakov, E. Maertz, R. H. T. Wilke, Qi Li, A. Soukiassian, D. G. Schlom, J. M. Redwing, A. Findikoglu, X. X. Xi, IEEE Trans. Appl. Supercond. 17 (2007) 2854.
[76] B. A. Glowacki, M. Majoros, M. E. Vickers, J. E. Evetts, Y. Shi, I. McDougall, Supercond. Sci. Technol. 14 (2001) 193.
[77] B. A. Glowacki, M. Majoros, M. E. Vickers, B. Zeimetz, Physica C 372 (2002) 1254.
[78] S. Soltanian, X. L. Wang, J. Horvat, A. H. Li, H. K. Liu, S. X. Dou, Physica C 382 (2000) 187.
[79] S. Zhou, A. V. Pan, M. Ionescu, H. Liu, S. Dou, Supercond. Sci. Technol. 15 (2002) 236.
[80] E. Martinez, L. A. Angurel, R. Navarro, Supercond. Sci. Technol. 15 (2002) 1043.

[81] N. M. Strickland, R. G. Buckley, A. Otto, Appl. Phys. Lett. 83 (2003) 326.
[82] Z. Q. Ma, Y. C. Liu, Q. Z. Shi, Q. Zhao, Z. M. Gao, Supercond. Sci. Technol. 21 (2008) 065004.
[83] Q. W. Yao, X. L. Wang, S. Soltanian, A. H. Li, J. Horvat, S. X. Dou, Ceramics International 30 (2004) 1603.
[84] F. Li, T. Guo, K. Zhang, L-P Chen, C. Chen, Q. Feng, Supercond. Sci. Technol. 19 (2006) 1196.
[85] A. Goyal et al., Appl. Phys. Lett. 69 (1996) 1795.
[86] E. Samuelsson, A. Mitchell, Metall. Trans. B 23B (1992) 805.
[87] Mahipal Ranot, W. K. Seong, Soon-Gil Jung, W. N. Kang, J. Korean Phys. Soc. 54 (2009) 2343.
[88] T. G. Lee, M. Ranot, W. K. Seong, S.-G. Jung, W. N. Kang, J. H. Joo, C.-J. Kim, B.-H. Jun, Y. Kim, Y. Zhao, S. X. Dou, Supercond. Sci. Technol. 22 (2009) 045006.
[89] C. J. Olson, F. Nori, Physica C 363 (2001) 67.
[90] M. Y. Kameneva, A. I. Romanenko, O. B. Anikeeva, N. V. Kuratieva, Physica C 408 (2004) 816.
[91] Y. Bugoslavsky, L. F. Cohen, G. K. Perkins, M. Polichetti, T. J. Tate, R. Gwilliam, A. D. Caplin, Nature 411 (2001) 561.
[92] M. Putti et al., Appl. Phys. Lett. 86 (2005) 112503.
[93] I. Pallecchi et al., Phys. Rev. B 71 (2005) 212507.
[94] E. M. Choi, H. S. Lee, H. Kim, S. I. Lee, H. J. Kim, W. N. Kang, Appl. Phys. Lett. 84 (2004) 82.
[95] Y. Zhao, C. H. Cheng, L. Zhou, Y. Wu, T. Machi, Y. Fudamoto, N. Koshizuka, M. Murakami, Appl. Phys. Lett. 79 (2001) 1154.
[96] S. F. Wang, S. Y. Dai, Y. L. Zhou, Y. B. Zhu, Z. H. Chen, H. B. Lü, G. Z. Yang, J. Supercond. 17 (2004) 397.
[97] T. Takenobu, T. Ito, D. H. Chi, K. Prassides, Y. Iwasa, Phys. Rev. B 64 (2001) 134513.
[98] S. X. Dou, S. Soltanian, J. Horvat, X. L. Wang, S. H. Zhou, M. Ionescu, H. K. Liu, P. Munroe, M. Tomsic, Appl. Phys. Lett. 81 (2002) 3419.
[99] K. S. Tan, S. K. Chen, B-H Jun and C-J Kim, Supercond. Sci. Technol. 21 (2008) 105013.
[100] A.V. Pogrebnyakov et al., Appl. Phys. Lett. 85 (2004) 2017.
[101] C. G. Zhuang, S. Meng, H. Yang, Y. Jia, H. H. Wen, X. X. Xi, Q. R. Feng, Z. Z. Gan, Supercond. Sci. Technol. 21 (2008) 082002.
[102] M. Ranot, W. K. Seong, S.–G. Jung, N. H. Lee, W. N. Kang, J. H. Joo, Y. Zhao, S.X. Dou, Physica C 469 (2009) 1563.
[103] M. Ranot, Soon-Gil Jung, W. K. Seong, N. H. Lee, W. N. Kang, J. Joo, C.–J. Kim, B.–H. Jun, S. Oh, Physica C 470 (2010) S1000.
[104] T. J. Jackson, S. B. Palmer, J. Phys. D: Appl. Phys. 27 (1994) 1581.
[105] S.-G. Jung, N. H. Lee, W. K. Seong, W. N. Kang, E.-M. Choi, S.-I. Lee, Supercond. Sci. Technol. 21 (2008) 085017.
[106] W. X. Li, R. Zeng, L. Lu, Y. Zhang, S. X. Dou, Y. Li, R. H. Chen, M. Y. Zhu, Physica C 469 (2009) 1519.
[107] W. X. Li, R. Zeng, L. Lu, S. X. Dou, J. Appl. Phys. 109 (2011) 07E108.
[108] S. X. Dou et al., Phys. Rev. Lett. 98 (2007) 097002.
[109] J. L. Wang, R. Zeng, J. H. Kim, L. Lu, S. X. Dou, Phys. Rev. B 77 (2008) 174501.
[110] A. Berenov et al., J. Mater. Res. 18 (2003) 956.
[111] S.-F. Wang, Z. Liu, Y.-L. Zhou, Y.-B. Zhu, Z.-H. Chen, H.-B. Lu, B.-L. Cheng, G.-Z. Yang, Supercond. Sci. Technol. 17 (2004) 1126.
[112] M. Hanna, S. Wang, J. M. Redwing, X. X. Xi, K. Salama, Supercond. Sci. Technol. 22 (2009) 015024.
[113] M. Ranot, K. Cho, W. K. Seong, S. Oh, K. C. Chung, W. N. Kang, Physica C 471 (2011) 582.
[114] A. V. Pogrebnyakov et al., Phys. Rev. Lett. 93 (2004) 147006.
[115] A. V. Pogrebnyakov, J. M. Redwing, J. E. Jones, X. X. Xi, S. Y. Xu, Qi Li, V. Vaithyanathan, D. G. Schlom, Appl. Phys. Lett. 82 (2003) 4319.
[116] S. R. Foltyn, Q. X. Jia, P. N. Arendt, L. Kinder, Y. Fan, J. F. Smith, Appl. Phys. Lett. 75 (1999) 3692.